\documentclass[%
 reprint,
 amsmath,
 amssymb,
 aps,
floatfix,
]{revtex4-2}

\usepackage{graphicx}
\usepackage{dcolumn}
\usepackage{bm}
\usepackage[dvipsnames]{xcolor}

\begin{document}

\preprint{APS/123-QED}

\title{Characterising a tunable, pulsed atomic beam using matter-wave interferometry}

\author{J Morley}
\author{R. Flack}%
\author{B. J. Hiley}%
\author{P. F. Barker}%
 \email{p.barker@ucl.ac.uk}
\affiliation{%
 Department of Physics and Astronomy,\\ University College London, London, UK
}%

\date{\today}

\begin{abstract}
We describe the creation and characterisation of a velocity tunable, spin-polarized beam of slow metastable argon atoms. We show that beam velocity can be determined with a precision below 1 \%  using matter-wave interferometry. The profile of the interference pattern was also used to determine the velocity spread of the beam, as well as Van der Waals co-efficient for the interaction between the metastable atoms and the multi-slit silicon nitride grating.  The Van der Waals co-efficient was determined to be $C_3$=1.84$\pm$0.17\,a.u., in good agreement with values derived from spectroscopic data. Finally, the spin polarization of the beam produced during acceleration of the beam was also measured, demonstrating a spatially uniform spin polarization of 96 \% in the m=+2 state. \end{abstract}
\maketitle
\section{Introduction}
Atomic and molecular beams are vitally important tools in physics, chemistry and materials science. They allow the study of both short and long range atomic and molecular interactions \cite{ramseybeams1} and, are an enabling technology for scattering experiments that provide elastic and inelastic cross-sections for scientific \cite{PhysRevLett.125.263401}, astrophysical \cite{O_Connor_2015} and industrial processes \cite{CHO1975157}. Experiments often require a well-defined velocity and velocity spread which is conventionally engineered through control of gas temperature, pressure and flow geometry\cite{beams1}. More recently, optical forces have been used to create tunable, low energy beams\cite{Cacciapuoti2001,Wang2003,barker,Lu2008,Taillandier-Loize2016a}, allowing much lower velocities with additional control over the beam velocity and velocity spread. Such beams are particularly useful for matter-wave interferometry and for exploring low energy atomic and molecular collisions\cite{Rakonjac:12}. 

An important requirement for quantitative analysis of many experiments with atomic beams is an accurate knowledge of their velocity/energy and also the spread in these values. Time-of-flight (TOF) methods, which determine an average atomic velocity, are commonly used and are simple to implement. However, for beams with time dependent accelerating forces, such as in optical acceleration, TOF calculations often do not accurately estimate the atomic velocity. To address this problem, alternative methods that measure the atomic velocity at a particular location have been employed. Optical techniques such as Ramsey interferometry\cite{Bateman2018} and Doppler shifted absorption spectroscopy \cite{Slowe2005} has been demonstrated for measuring ultra low velocity atomic beams.

Matter-wave interferometry has developed considerably since its first demonstration using electrons by Davison and Germer in 1927\cite{davisson}. For example, interferometry with cold atoms and molecules have been used to explore the mass limits to quantum superposition\cite{arndt}, for measuring the gravitational constant\cite{tino}. It has also been used for metrological applications including gravimetry\cite{Peters_2001}, the measurement of polarisability \cite{PhysRevA.76.013607} and for absolution absorption measurements\cite{PhysRevA.78.063607}.  In this paper, we use matter-wave interferometry through a multi-slit grating to characterise a slow atomic beam of metstable argon atoms. In particular, we use it determine the velocity of the tunable atomic beam, with significantly higher precision than time-of-flight techniques. Additionally, using the same method allows us to determine the $C_3$ coefficient that characterises the Van der Waals interaction between these atoms and the silicon nitride diffraction grating.

\section{Matter-wave interferometry with a nanomechanical grating }
 To undertake these experiments, we utilise a multi-slit matter-wave interferometer, with a detector that allows us to record the interference pattern, as well as the time-of-flight of a packets of atoms within the pulsed beam. The shape and fringe spacing of the interferogram is dependent on the de Broglie wavelength, $\lambda_{DB}=h/mv$, of the atoms in the beam and a measurement of $\lambda_{DB}$ allows us to determine the velocity, $v$, where $h$ is Planck's constant and $m$ the atomic mass.
 
For a plane wave of wavenumber $k$, and momentum, $\hbar k=\frac{2\pi}{\lambda_{DB}}$, incident on a grating with $N$ slits, the signal intensity $I_0$, as a function of velocity and transverse position $x$ observed on the detector, is a product of the incident intensity $I_{inc}$, the grating function, and a slit function $f_{slit}$ as\cite{Grisenti1999},
\begin{equation} \label{eq:1}
I_0(x,v)=I_{inc}\left[\frac{\sin\left(\frac{1}{2}N k v d\,\sin(\frac{x}{L})\right)}{\sin\left(\frac{1}{2}k v d\,\sin(\frac{x}{L})\right)}\right]^2|f_{slit}(x,v)|^2,
\end{equation}
where $L$ is the grating-detector distance and $d$ is the grating slit separation. Both functions are modified by the atomic velocity $v$ and so can be used to characterise the velocity of the beam. 
The slit function is given by
\begin{equation} \label{eq:2}
f_{slit}(x,v)=\frac{2\cos(\frac{x}{L})}{\sqrt{\lambda_{DB}(v)}}\int^{\frac{s_0}{2}}_0\,d\zeta
\cos\left[k v\,\sin\left(\frac{x}{L}\right)\left(\frac{s_0}{2}-\zeta\right)\right]\tau(\zeta,v),
\end{equation}
in which the standard single slit diffraction function is modified by $\tau$, which accounts for the attractive Van der Waals (VdW) potential, $V=-C_3/l^3$, between the atoms and the slit walls separated by $l$. The actual slit width is $s_0$,  $\zeta=s_0/2-x$ and $\tau$ is given by
\begin{equation} \label{eq:3}
\tau(\zeta,v) = exp\left[i\frac{t \cos\beta}{\hbar v}\frac{C_3}{\zeta^3}\frac{1+\frac{t}{2\zeta}\tan\beta}{(1+\frac{t}{\zeta}\tan\beta)^2}\right],
\end{equation}
where $C_3$ is the VdW coefficient, $t$ is the grating thickness and $\beta$ the angle of the tapered edge of the slit profile.

\begin{figure}[hbt!]
    \includegraphics[width=80mm]{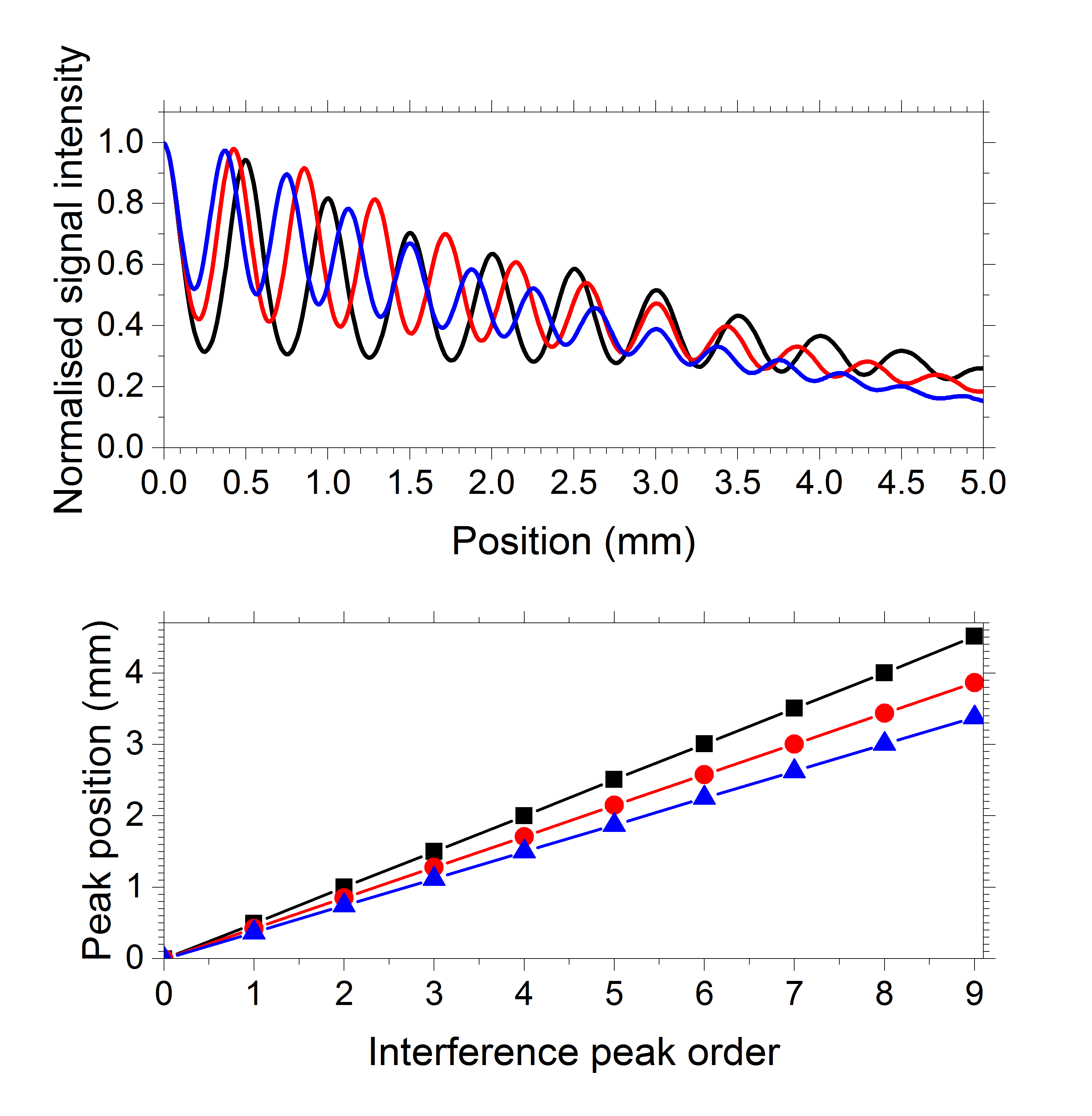}
    \caption{(a) Equation \ref{eq:1} across 5\,mm showing one half of the symmetric interference signal. As the atomic velocity changes from 12\,ms$^{-1}$ (black), 14\,ms$^{-1}$ (red) and 16\,ms$^{-1}$ (blue) the peak positions shift. In each case the velocity spread is 1\,ms$^{-1}$.
    \newline (b) The change in peak positions of successive orders for 12\,ms$^{-1}$ (black), 14\,ms$^{-1}$ (red) and 16\,ms$^{-1}$ (blue). This has been calculated for a velocity distribution of 1\,ms$^{-1}$ and $C_3$ value of 2.17\,a.u.}
    \label{CompareV}
 \end{figure}

The transverse coherence of the atomic beam contributes to the contrast of the fringes and is included in the model by convolving equation \ref{eq:3} with the atomic beam profile, $A(x)$. The profile, which can be seen in figure \ref{SignalProfileRel}, is defined by the collimation slit configuration, scattering from the slit walls and the velocity group of the atoms being measured. This is discussed in further detail below. The beam width also defines the number of grating slits that are being illuminated, which leads to a new intensity given as
\begin{equation} \label{eq:4}
I_{c}(x,v) =\int_{-\infty}^{\infty} I_0(x',v) A(x-x') dx.
\end{equation}
Finally, to account for the observed spread of longitudinal velocities, equation \ref{eq:4} is integrated over the longitudinal velocity spread.
\begin{equation} \label{eq:6}
I(x,v) =\int_{-\Delta v/2}^{\Delta v/2} I_c(x,v) dv.
\end{equation}
This intensity profile can be fitted to the experimental data to measure $v,\Delta v,$ and also $C_3$. An example of how the peak velocity modifies the pattern is shown in figure \ref{CompareV}(a). To provide an initial estimate of atomic velocity, $v$, the position of each interference order is plotted for the three atomic velocities shown in figure \ref{CompareV}(b). The gradient of the fitted line is equal to $\frac{h}{mv} \frac{L}{d}$.
 
\begin{figure}
    \includegraphics[width=80mm]{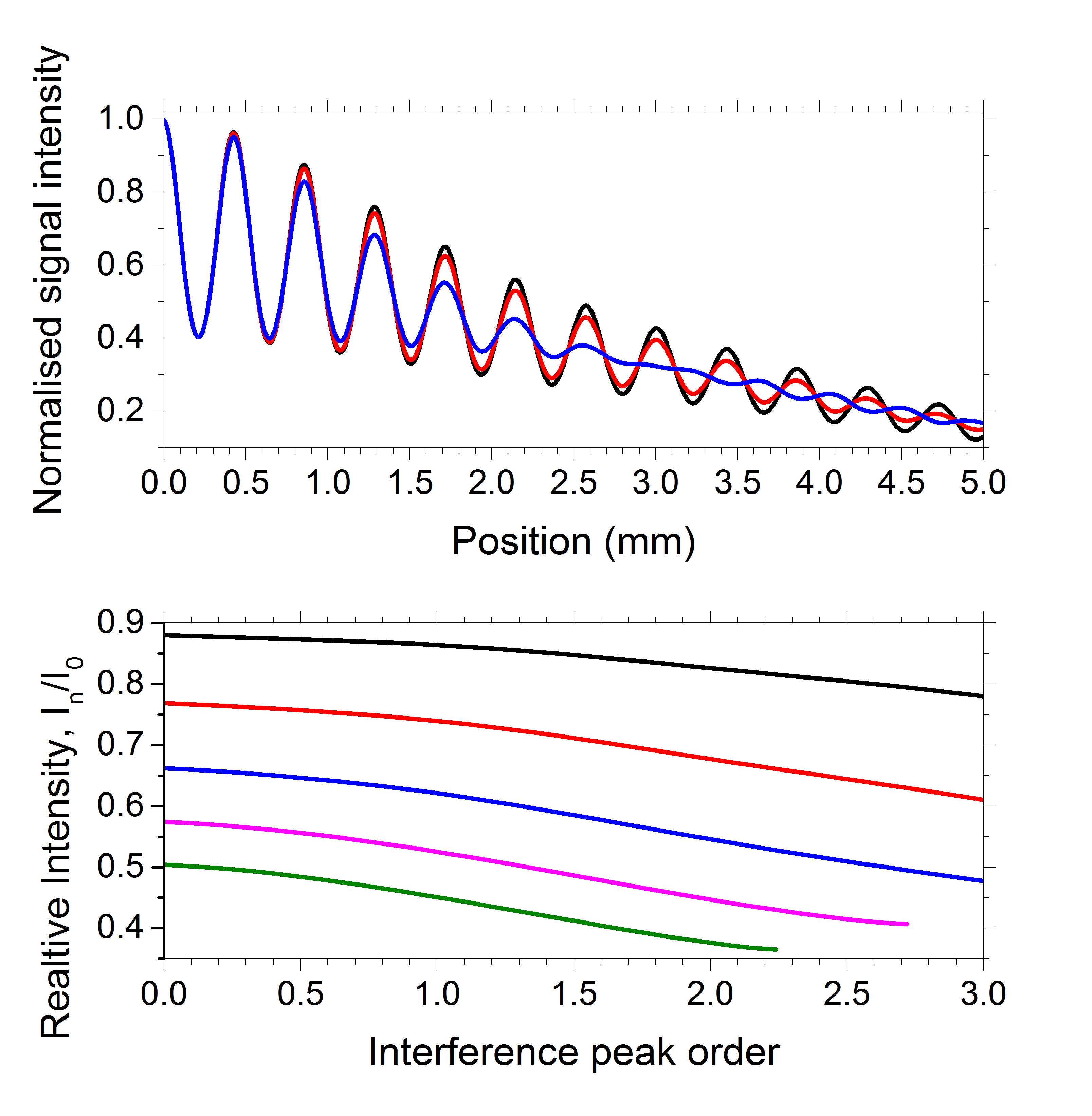}
    \caption{(a)The change in the interference pattern for a peak velocity of 14\,ms$^{-1}$, as the velocity distribution changes from 0.5\,ms$^{-1}$ (black), 1\,ms$^{-1}$ (red) and 2\,ms$^{-1}$ (blue). The $C_3$ coefficient is 1.55\,a.u..
    \newline (b) The relative peak height, measured from the 0th order peak height for the 2nd (black), 3rd (red), 4th (blue) and 5th (pink) interference orders. Plotted for a peak velocity of 14\,ms$^{-1}$ and $C_3$ coefficient of 1.55\,a.u.}
    \label{dVComb}
 \end{figure}

Although, the velocity spread affects the peak height and contrast, and to some extent the fringe spacing shown in figure \ref{dVComb}(a), to a good approximation, a measurement of the fringe spacing of the interferograms gives a very good estimate of the velocity. The modification due to the velocity spread described in equation \ref{eq:6} is more pronounced at higher interference orders, and therefore, the shape of the interferogram can be used to characterise the beam's velocity spread and the effective grating slit width.
 A measurement of the peak contrast is isolated from the peak height in figure \ref{dVComb}(b) since the peak height is also dependent upon the peak velocity and the VdW interaction of the slit function.
 
\begin{figure}
    \centering
    \includegraphics[width=80mm]{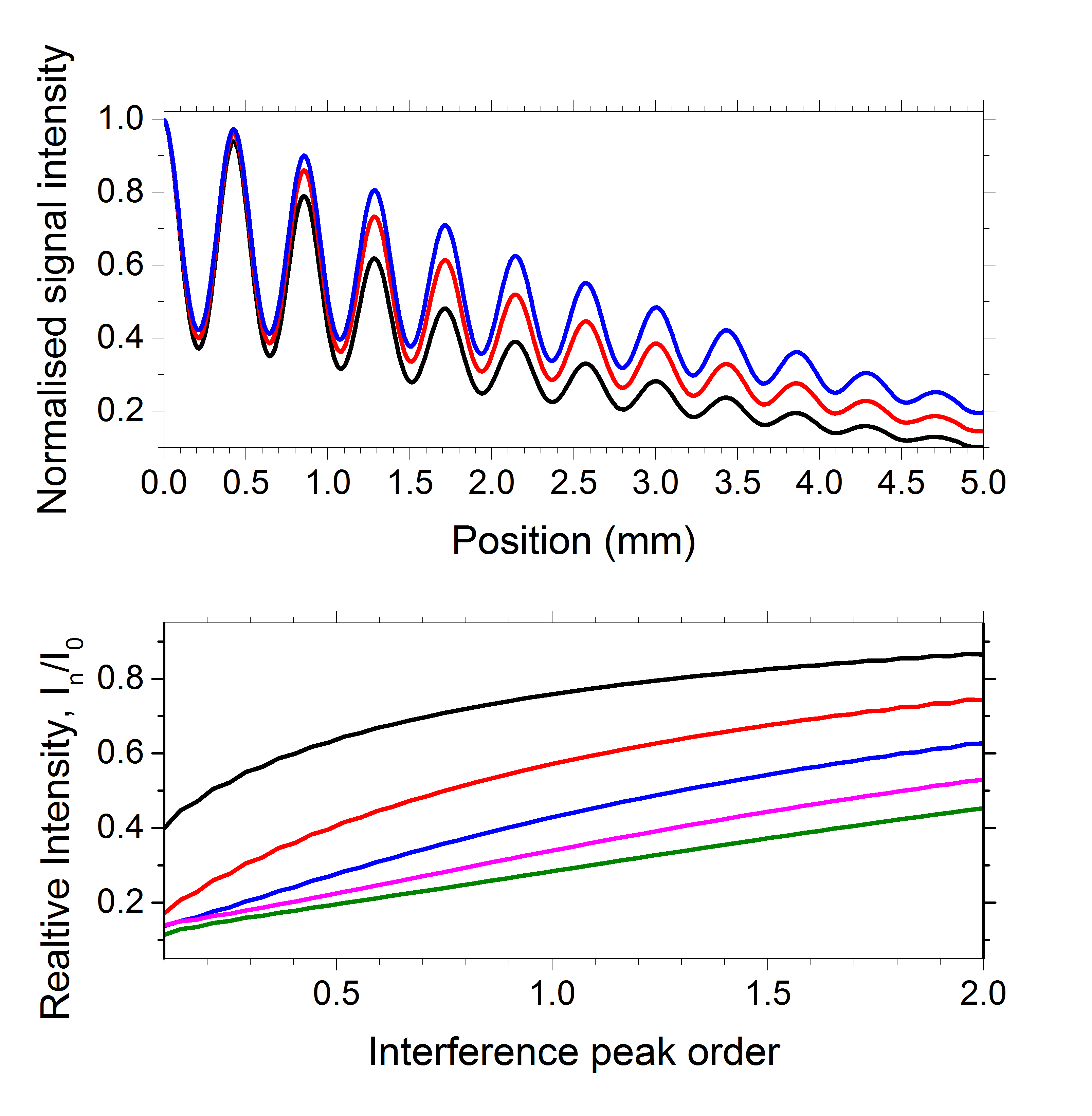}
   \caption{(a) Interferograms with a peak atomic velocity of 14\,ms$^{-1}$ and velocity spread of 1\,ms$^{-1}$ for a VdW coefficient of 1\,a.u. (black), 1.5\,a.u. (red), 2\,a.u. (blue).
   \newline (b) The relative peak height, measured from the 0th order peak height plotted over a changing of VdW potential, for the 2nd (black), 3rd (red), 4th (blue) and 5th (pink) interference order fringe. For an atomic velocity of 14\,ms$^{-1}$ and velocity spread of 1\,ms$^{-1}$}
   \label{C3Comb}
 \end{figure}

With the average velocity measured directly from the fringe spacing, the upper envelope of the signal contains the information to measure $\Delta v$ and $C_3$. The VdW potential changes the slit function and hence the upper envelope of the signal, which simultaneously adjusts the peak height and contrast as shown in figure \ref{C3Comb}(a). As a result, $\Delta v$ and $C_3$ are varied when fitting equations 1 to 5 and have co-dependent uncertainties. In figure \ref{C3Comb}(b), the height of the 2nd-6th order of interference peaks are plotted for a changing VdW potential. 
 
\section{Tunable, low-velocity, metastable argon beam}
 We create the tunable, low-energy, source of metastable argon by accelerating them out of a magneto optical trap (MOT) using a single laser beam as shown in figure \ref{Schematic}. To load the MOT, a thermal effusive beam of metastable argon atoms is decelerated in a Zeeman slower. Neutral atoms are introduced into an RF discharge by the effusive beam of neutral argon atoms\cite{edmunds}. The MOT sits in a chamber held at 1$\times$10$^{-8}$\,mbar and the atoms are laser cooled using 3 orthogonal, counter propagating beams arranged such that there is no cooling beam in the vertical axis. We expand upon the work of Tailandire et al.\cite{Taillandier-Loize2016a} by selecting the transverse velocity, as well as the longitudinal velocity. This makes the transverse velocity dependent on fixed geometry rather than only the temperature of the trapped atoms. Additionally, a larger, cooler MOT can be obtained, and could in theory further reduce the atomic beam's lower velocity limit.

 \begin{figure}[htb!]
    \centering
    \includegraphics[width=100mm]{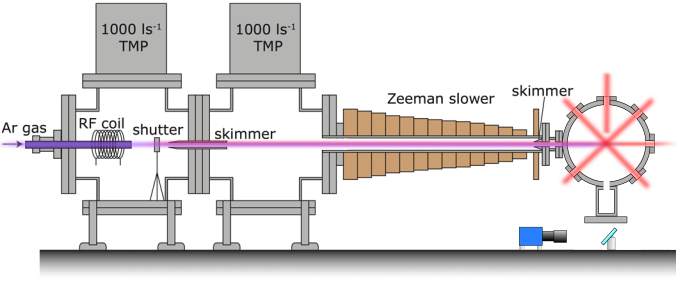}
    \caption{The experimental arrangement of the metastable argon MOT, the pulsed atomic beam and the inteferometer. Two, 200CF, 6-way cross chambers house effusive atomic beam, including a skimmer between the chambers to collimate the beam. After the 2nd chamber, the Zeeman slower coils extend towards the spherical octagon MOT chamber, while the Zeeman beam counter-propagates to the atomic beam. The MOT chamber contains the trapping and push lasers. Attached to the bottom of this chamber is the detector which records the matter-wave interferograms and the time-of-flight of the atoms. This signal is recorded on a CMOS camera.}
    \label{Vacuum}
 \end{figure}

The cold atomic cloud within the MOT has a 1/e$^2$ Gaussian radius of 0.5 mm and contains up to 3x10$^6$ atoms in the metastable 4s[3/2]\textsubscript{2} state.  A maximum density of approximately 10$^{10}$ cm$^{-1}$ is determined by Penning and associative ionization. To create the low velocity, pulse atomic beam, the atoms are accumulated in the MOT for 500\,ms seconds before a short pulse of between 0.6 and 7 ms, from a near resonant laser, illuminates the atomic cloud. This pushes the atoms downwards via the radiation pressure force in the direction of the vertical push beam, due to repeated recoil events following the absorption and emission of push beam photons. The push beam passes through an AOM allowing both the fast switching needed for sub millisecond pulse widths and for the tuning of the push beam frequency. After acceleration from the optical forces, the atoms pass through collimating slits placed 13\,mm and 43\,mm below the cold atomic cloud. The accelerated atoms travel downwards to a microchannel plate detector with phosphor screen stack. This is placed approximately 160\,mm below the slits. When neutral atoms in the metastable state strike the surface of the detector they are ionized and create a shower of electrons which are detected on the phosphor screen. The phosphor screen is imaged through a viewport using a CCD camera. A gated voltage can be applied to the MCP detector allows imaging of the beam profile and interference patterns with a temporal resolution down to 0.1 $\mu$s.  For characterising the beam properties, a silicon nitride grating with a slit spacing of 257\,nm, slit width 90\,nm and a depth of 100\,nm is placed just below the last collimating slit. 
 \begin{figure}[htb]
    \centering
    \includegraphics[width=70mm]{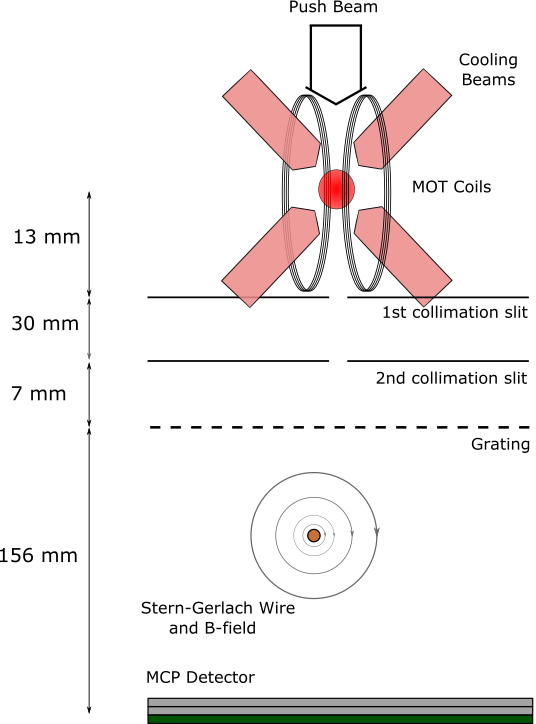}
    \caption{A schematic showing the layout of the experiment, including both the matter wave interferometer and a Stern-Gerlach wire which provides a B-field gradient to separate the 5 magnetic sublevels in the beam. }
    \label{Schematic}
 \end{figure}
 
  \subsection{Longitudinal Velocity}
\begin{figure}
    \centering
    \includegraphics[width=80mm]{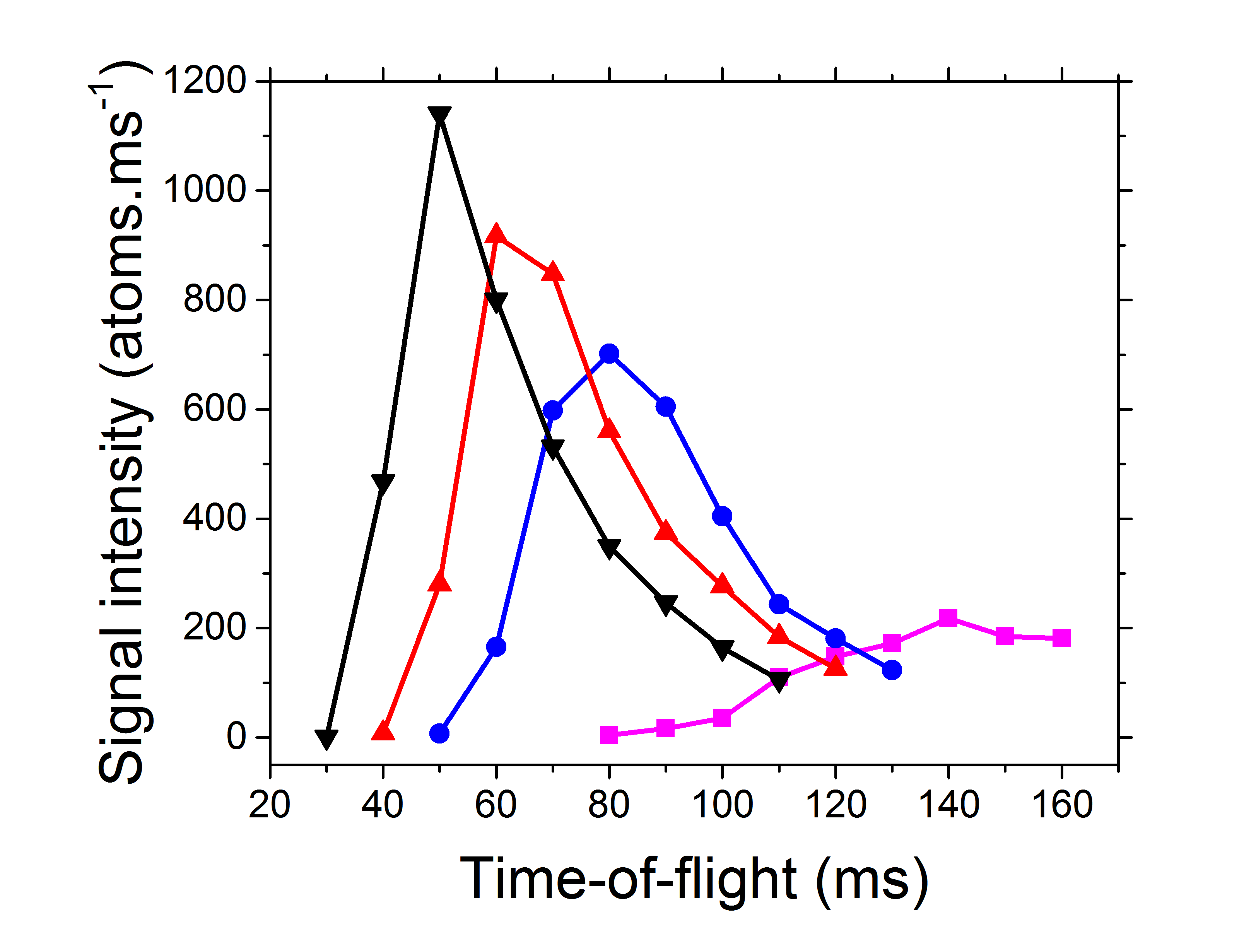}
    \caption{The distribution of the atom beam's TOF for push beam pulse lengths of 7\,ms (black), 4\,ms (red), 5\,ms (blue) and 2 ms (magenta). Such long pulse lengths and TOF are possible with low push beam power and with no collimation slits.}
    \label{PBTimeVel}
 \end{figure}

 The longitudinal velocity of the atomic beam can be varied between 10-60\,ms$^{-1}$ by adjusting the push beam pulse length, the beam intensity or its frequency detuning with respect to the atomic resonance. Increasing the pulse length increases the interaction time between the push beam and hence the period of acceleration. This effect is most clearly seen in figure \ref{PBTimeVel} which shows the TOF distribution for four different push beam pulse lengths, for a push beam detuned 15\,MHz from resonance.
 
\begin{figure}
    \centering
    \includegraphics[width=80mm]{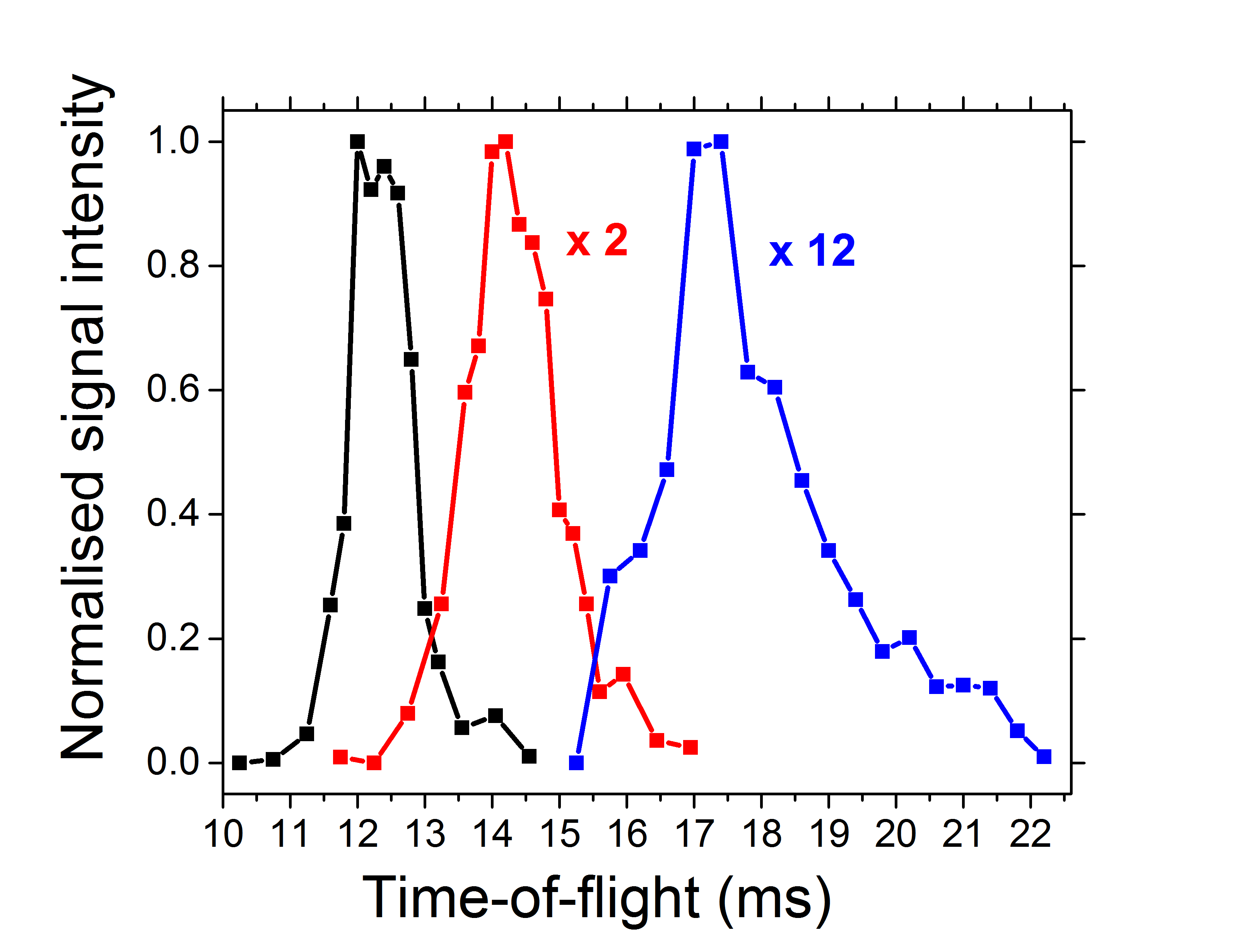}
    \caption{The effect of the frequency of the push beam pulse on the velocity of the atomic beam. For a pulse length of 0.6\,$\mu$s and detuned by frequency 4.76\,MHz (blue), 8.01\,MHz (red) and 11.31\,MHz (black). The beam intensity decreases with the velocity, so the signals are amplified as indicated.}
    \label{PBDetTOF}
 \end{figure}

 The push beam interaction time is ultimately limited by the attenuation from the 10\,$\mu$m collimation slit that is placed 12mm below the source. This causes the push beam to diverge rapidly after diffracting through the slit. This causes the push beam intensity to reduce by a factor of approximately 0.5$I_0$ for every mm below the slit which quickly reduces it effect. 
 Given that the atoms are accelerated through a spatially varying magnetic field (MOT coils shown in figure \ref{Schematic}), the atom's resonant transition frequency is shifted. The maximum force occurs when the laser is at resonance in 0\,T magnetic field. Maximal force can be applied by using a detuned the push beam to compensate for the Zeeman shift in the atomic resonance due to the magnetic field. The effect of the push beam frequency is shown in figure \ref{PBDetTOF} for a pulse length of 0.6\,ms.

\begin{figure}
    \centering
    \includegraphics[width=80mm]{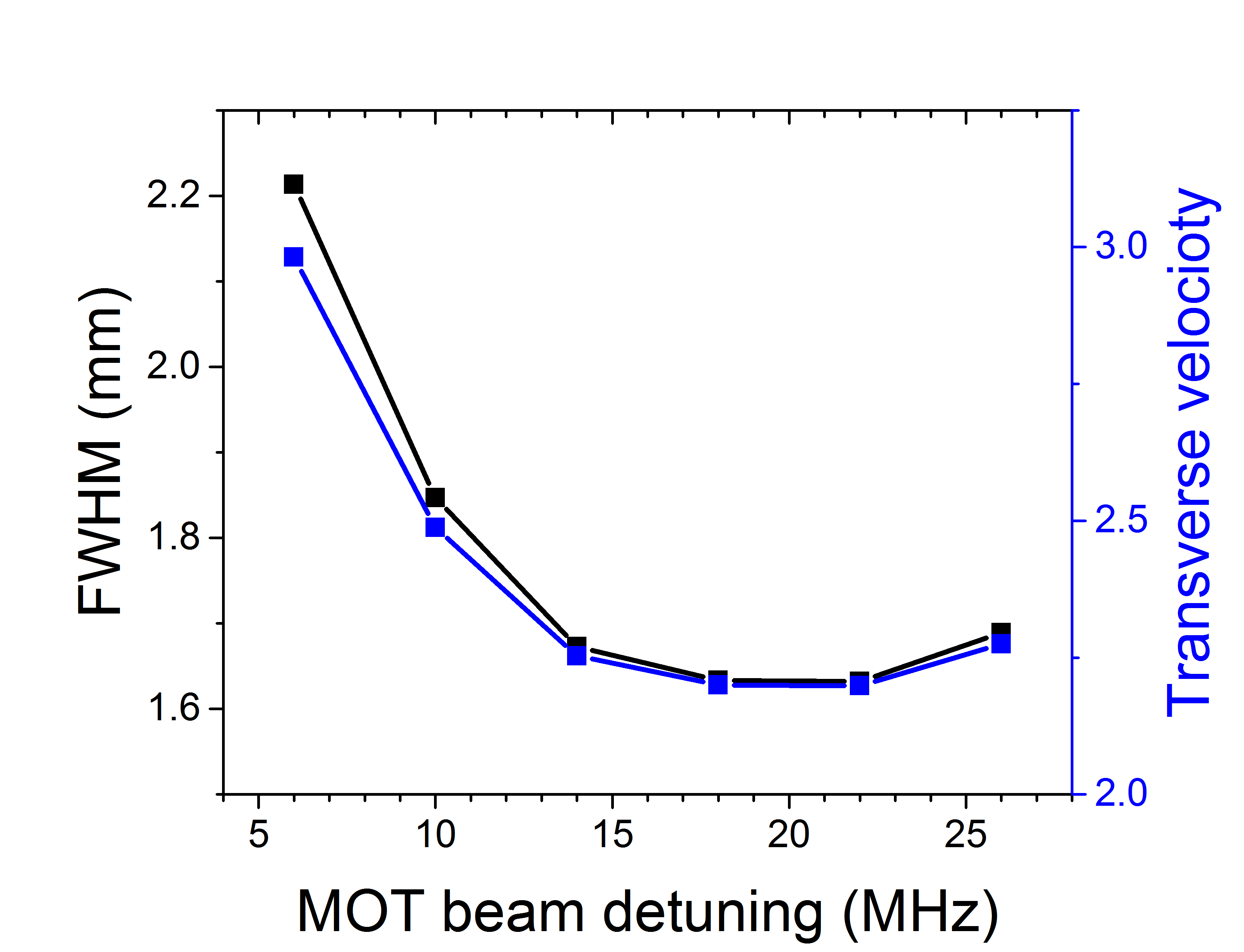}
    \caption{The divergence of the atomic beam (black squares) and the equivalent transverse velocity (blue squares) as a function of the MOT cooling beam detuning frequency, which controls the MOT temperature. Data is for a 4\,ms peak TOF, using a single 10$\mu$ collimation slit 13mm below the MOT.}
    \label{MOTDetDivVx}
 \end{figure}

 \begin{center}
\subsection{Transverse Velocity}
 \end{center} 
 With no collimation from the slits, the divergence angle of the atomic beam would only be dependent on the temperature of the MOT for a given longitudinal velocity. For an atomic beam with a 4.5\,$\pm$0.5\,ms$^{-1}$ TOF, a beam width of 4\,mm FWHM at the detector gives the beam's divergence angle of 19.4\,mrad and an average transverse atomic velocity of 0.45\,$\pm$0.05\,ms$^{-1}$. The MOT cooling beam frequency adjusts the temperature of the atoms in the MOT and changes the beam divergence, as seen in figure \ref{MOTDetDivVx}. The minimum beam width is limited by the minimum MOT temperature of approximately 100\,$\mu$K.
 
\begin{figure}
    \centering
    \includegraphics[width=80mm]{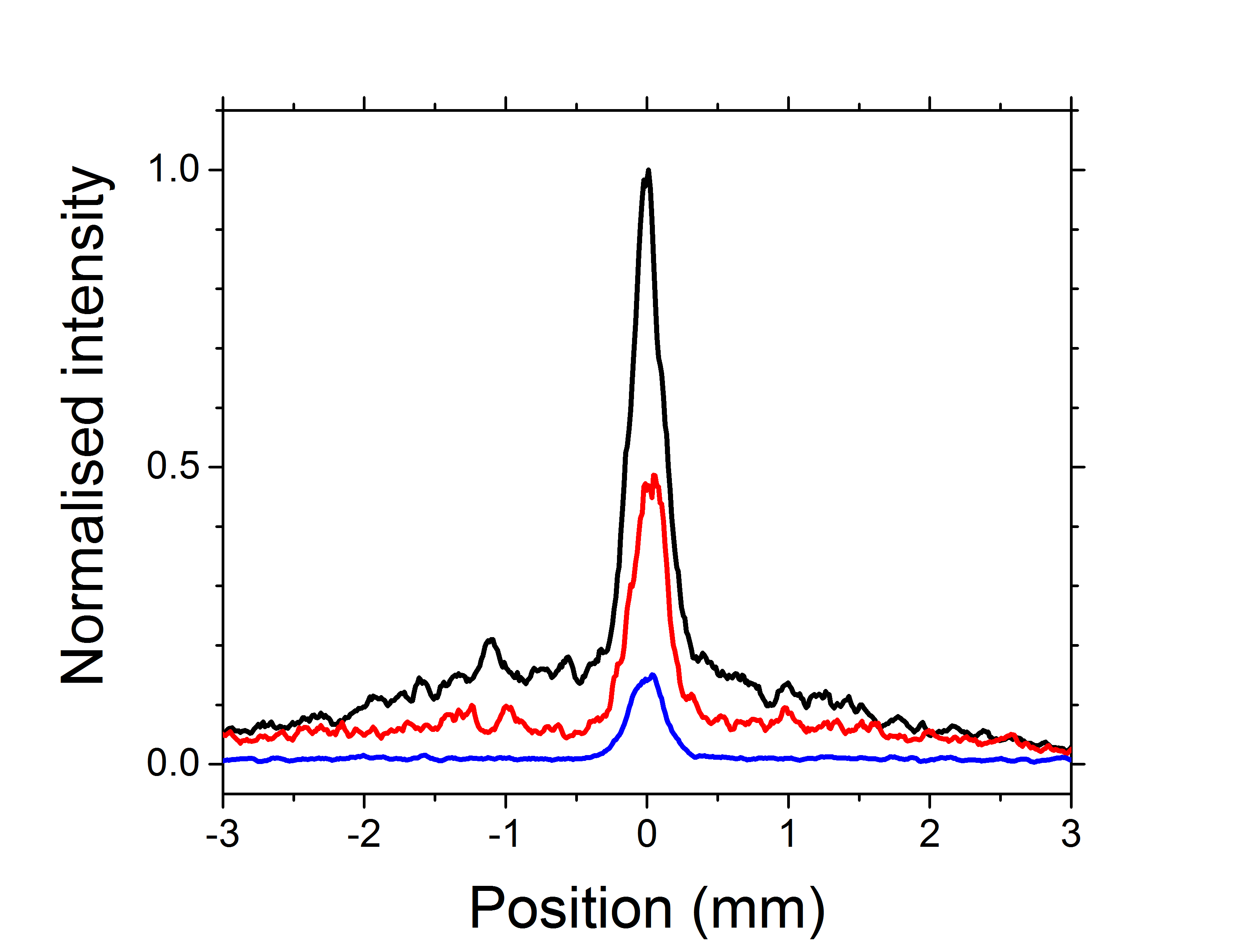}
    \caption{Measured atomic beam width profile for the 3 velocities that are used for the matter wave interferometry (approximately 11\,ms$^{-1}$ (blue), 15\,ms$^{-1}$ (red) and 18\,ms$^{-1}$ (black)). This beam width was narrowed using a 10\,$\mu$m slit and a 50\,$\mu$m for the initial and second collimating slit respectively. There is a background signal indicated by the lower, wider peak that arises due to elastic collisions with the walls of the slit.}
    \label{SignalProfileRel}
 \end{figure}

To observe interference fringes the beam must have sufficient transverse coherence. This is achieved by limiting the angle that the source subtends from the grating and limiting the spread of transverse velocity. To select the beam's transverse velocity independently of longitudinal velocity or MOT temperature, two collimation slits are placed between the MOT and the detector. The beam width is now controlled by the geometry of the collimation slit setup. With a 10\,$\mu$m slit, 13\,mm below the MOT and a 50\,$\mu$m slit, 43\,mm below the MOT, the beam divergence is 2.4\,mrad based on a FWHM of approximately 0.5\,mm  shown in figure \ref{SignalProfileRel}. Minimising the transverse velocity distribution increases the contrast of the matter-wave interference fringe peaks used to characterise the longitudinal velocity.
\subsection{Spin Polarisation of the Atomic Beam}
The 4s[3/2]\textsubscript{2} metastable state (J =2) has five m-states ($m=-2,-1,0,1,2$). For the push beam (and the cooling beams) the 4s[3/2]\textsubscript{2}-4p[5/2]\textsubscript{3} is closed cycle transition. If the push beam is right/left circularly polarised with respect to the atom's quantisation axis, the atoms can only be pumped to the m=-2 or m=-3 stretch state. 
 
\begin{figure}
    \centering
    \includegraphics[width=80mm]{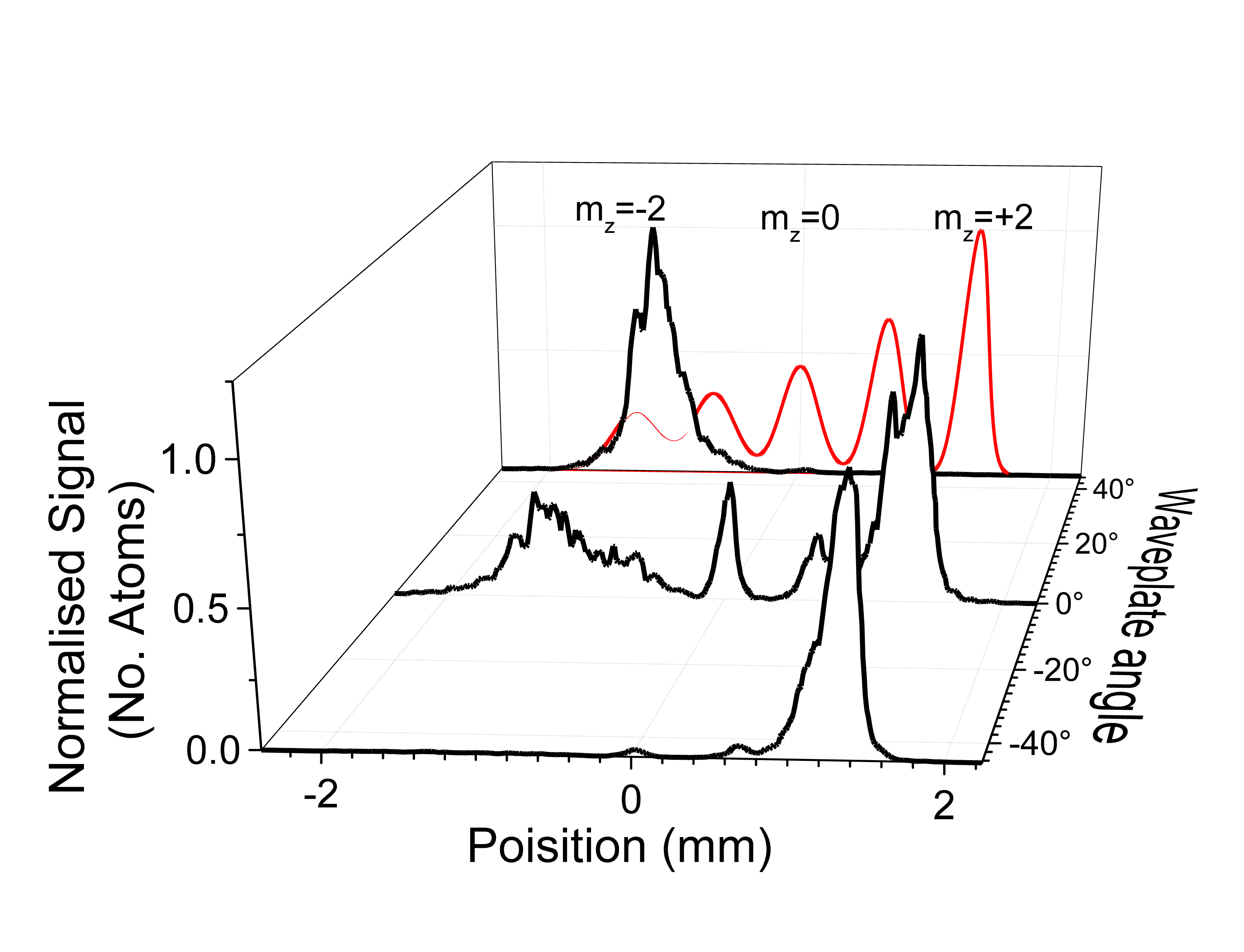}
    \caption{The $m$ state signal intensity as the push beam polarisation is changed from $\sigma^-$ through $\pi$ to $\sigma^+$ by adjusting the rotation of the $\frac{\lambda}{4}$ waveplate in the push beam path. The atomic beam passes by a Stern Gerlach wire which creates a magnetic field gradient to spatially separate the $m$ states.}
    \label{PBPol}
 \end{figure}

 To analyse the spin polarisation purity of the atomic beam, a Stern-Gerlach (SG) wire was placed between the second collimation slit and the detector shown in figure \ref{Schematic}. A current of around 200\,A was pulsed for up to 1\,ms through the wire at the moment the atoms pass the wire. The resultant magnetic field gradient spatially separates the $m$ states on the detector. Figure \ref{PBPol} shows the change in the atomic beam spin polarisation as the push beam polarisation is changed using a $\frac{\lambda}{4}$ waveplate. By finding the fractional area corresponding to m=+2 or m=-2 state, we determine that greater than 96\% of atoms can be placed in either of the stretch states during acceleration of the atoms with the push beam.

\section{Matter-Wave Interferometery of a Slow Metastable Argon Atomic Beam}
The collimated atomic beam is pulsed every 0.5\,s and interference fringes are observed on the MCP detector placed 156\,mm below the grating.

\begin{figure}
    \centering
    \includegraphics[width=75mm]{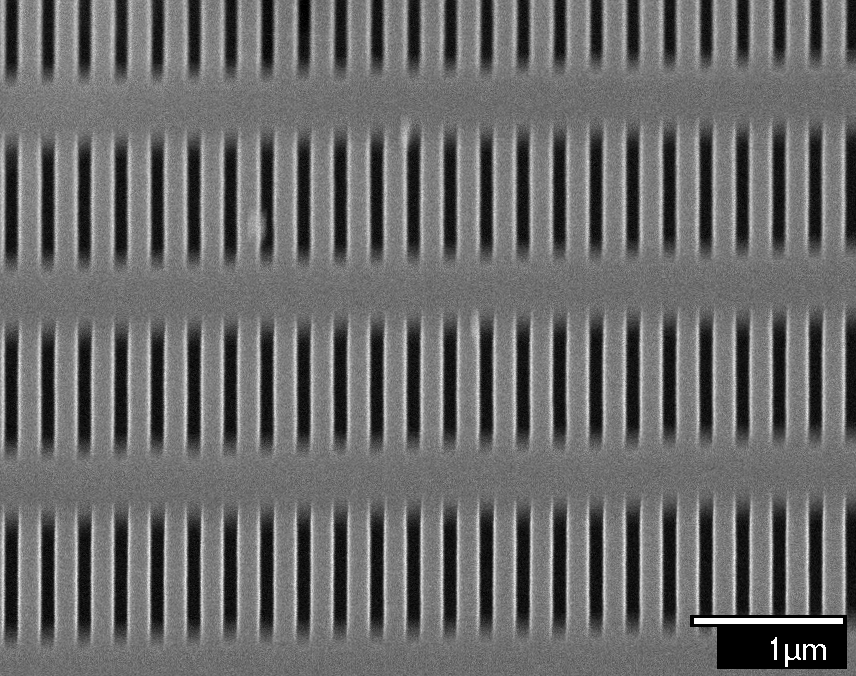}
    \caption{A SEM image of the 90x990\,nm grating slits created in a SiNi membrane. The dimensions are slightly distorted due to the non-perpendicular viewing angle.}
    \label{GratingWideScale}
 \end{figure}

The grating's slits lie in a 160\,nm thick SiN wafer, positioned 10\,mm below the second slit. The grating has a period of 257\,nm, with each slit 97\,$\mu$m wide, shown in figure \ref{GratingWideScale}. The atoms' final positions, after undergoing diffraction, are observed on the MCP detector. The MCPs have an ultimate spatial resolution of 6\,$\mu$m, but the reduced resolution of the phosphor screen increases this to $\approx$25\,$\mu$m. A 1392 x 1040 pixel CCD camera is focused on the screen, with each pixel covering 11.4\,$\mu$ m. Post processing of the images is used to find the centre of the Gaussian signal, and therefore the position of the atom, bringing the spatial resolution of the detector system down to 11.4\,$\mu$m.

The push beam is configured to produce the three velocity distributions shown in figure \ref{PBDetTOF}. Within each distribution, the detector is gated to select a narrow velocity range. For the three interferograms shown, the $v_{TOF}$ calculated to be 11.65\,$\pm$0.49\,ms$^{-1}$, 14.04\,$\pm$0.71\,ms$^{-1}$ and 16.67\,$\pm$0.54\,ms$^{-1}$ as shown in figures \ref{FringeFit} (a)-(c) respectively.

\begin{figure}
    \centering
    \includegraphics[width=70mm]{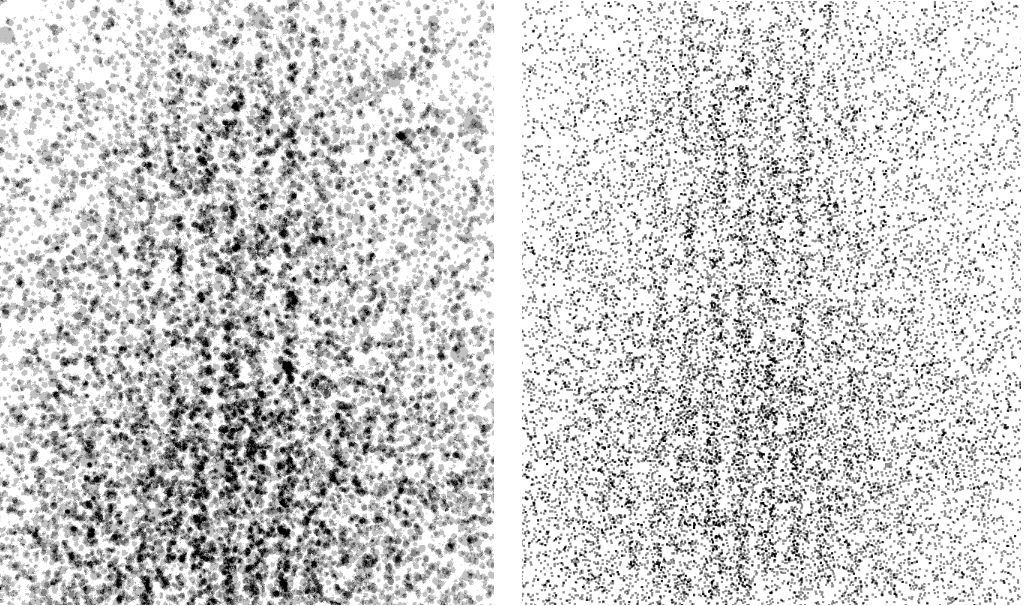}
    \caption{Detector images indicating the difference between the centroided (right) and raw (left) detector images for the $v_{TOF}$=11.65\,ms$^{-1}$ velocity atoms.}
    \label{Fringes}
 \end{figure}

\begin{figure}[htb]
    \centering
    \includegraphics[width=80mm]{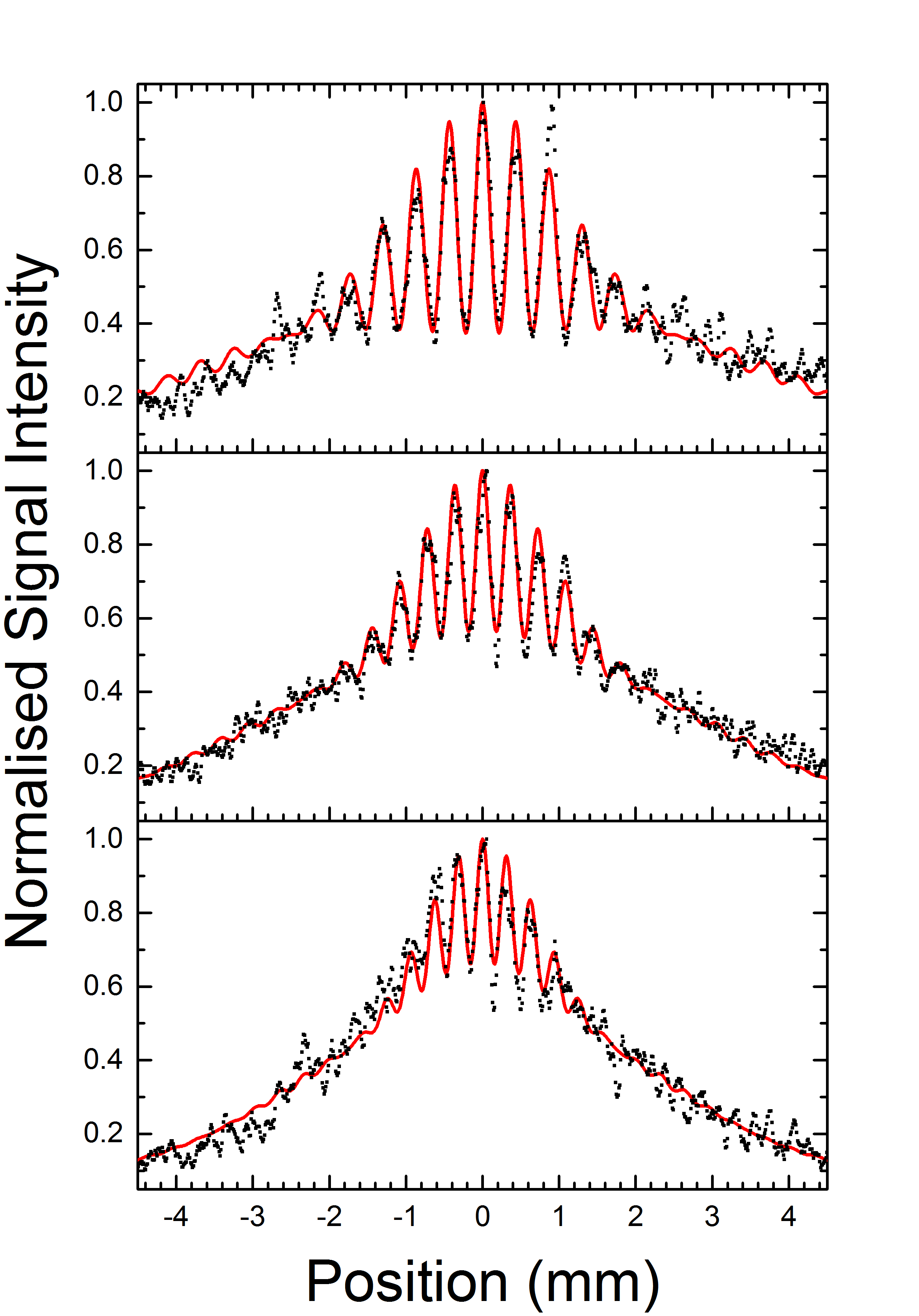}
    \caption{The interferograms for atoms with $v_{TOF}$ of (a) 11.65\,$\pm$0.49\,ms$^{-1}$, (b) 14.04\,$\pm$0.71\,ms$^{-1}$ and (c) 16.67\,$\pm$0.54\,ms$^{-1}$ (black squares). The red lines are fits to data using  \ref{eq:1}. The intensities have been normalised to the central peak.}
    \label{FringeFit}
 \end{figure}

The push beam configured for the atomic velocity used in figure \ref{FringeFit} (a) delivers an average of approximately 3 atoms per push cycle. By averaging the signal from 80,000 cycles, the interference fringes appear as shown in figure \ref{Fringes}.

 \subsection{Beam velocity from fringe spacing}
\begin{figure}
    \centering
    \includegraphics[width=80mm]{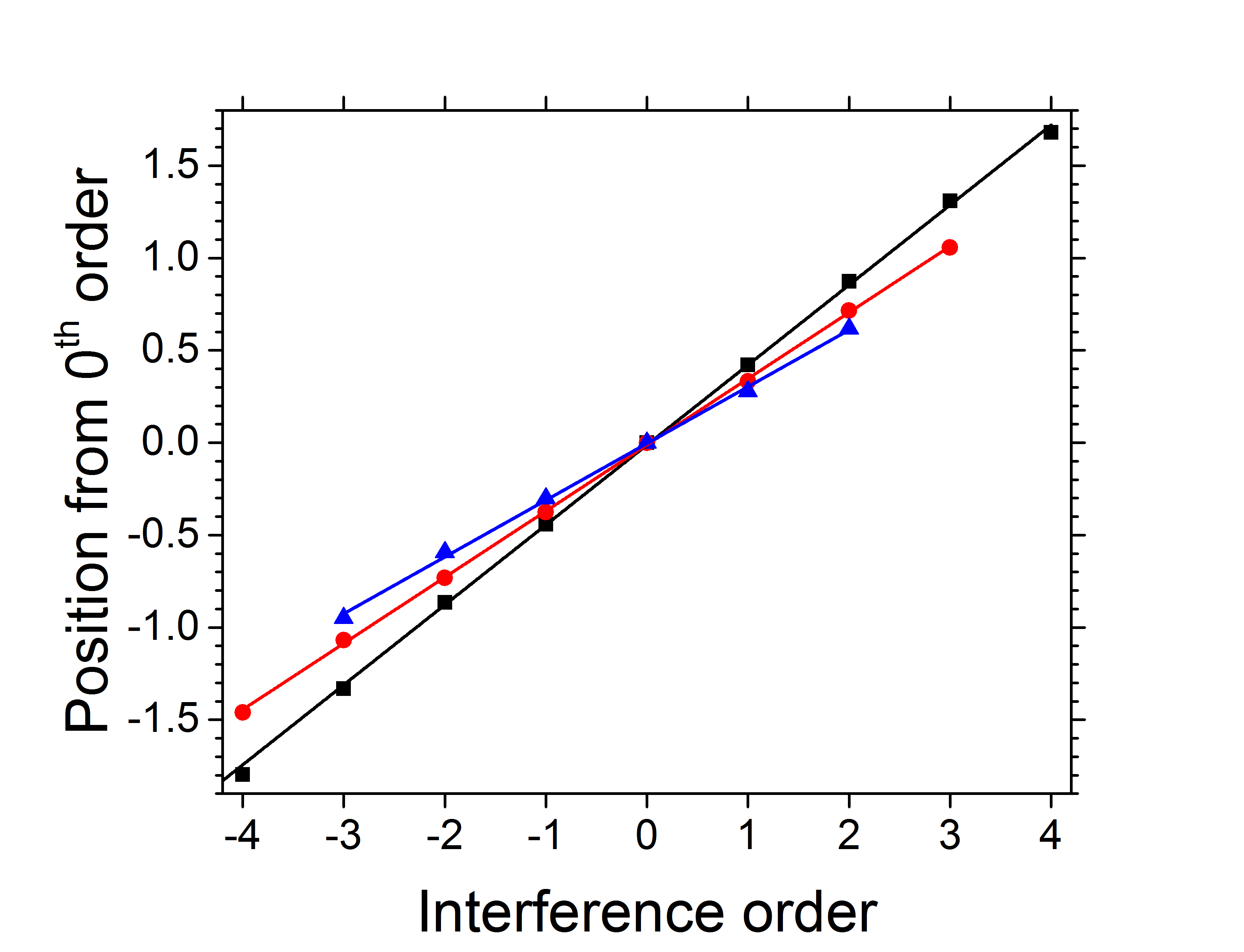}
    \caption{The interference peak positions in mm (with respect to the zeroth order peak) plotted as a function of the interference order for the three data sets in figure \ref{FringeFit}. The fitted line with gradient, $\frac{h}{mv} \frac{L}{d}$, provides a measurement of the velocity of 13.92$\pm$0.19 ms$^{-1}$ (black), 16.82$\pm$0.23 ms$^{-1}$ (red), 19.62$\pm$0.26 ms$^{-1}$ (blue).}
    \label{VMeasurement}
 \end{figure}

 The interference peak positions are measured for the three data sets and a linear fit for is shown for each in figure \ref{VMeasurement}. From the resulting gradient the average velocity was measured with these results as shown in table I.
 From this gradient the average velocity is measured to be 13.92\,$\pm$ 0.19\,ms$^{-1}$,  16.82\,$\pm$0.23\,\,ms$^{-1}$ and 19.62\,$\pm$0.26\,\,ms$^{-1}$ for the data sets from figures \ref{FringeFit}(a)-(c) respectively. 
 After the push beam interaction, the atoms travel towards the detector accelerated only by gravity. The increase in velocity due to gravity is $<$1\% of these values, so after the interaction with the grating, the atomic velocity is to a very good approximation constant. The values are calculated using $L$=156$\pm 1$\,mm and $d$=257$\pm$3\,nm, The grating spacing, $d$, is best measured using SEM imaging which can achieve an accuracy of up 0.4\,nm. However, typically these are between 1-20\,nm depending on the instrument and its calibration. The varying peak positions have an uncertainty of 0.3\%.
As the velocity spread increases, there is slight change in the higher order peak positions and there is subsequently a change in the fitted gradient. The effect causes a 0.1\,ms$^{-1}$ change in the measured velocity for every 1\,ms$^{-1}$ of spread in the atomic beam.
\subsection{Beam velocity from fitting}
To include the beam's velocity spread, $\Delta v$, and measure the velocity $v$ more precisely, equation \ref{eq:6} is fitted to the interferogram using a standard least squares method. The free parameters are $v, \Delta v$ and $C_3$ which gives $v_g=13.90\pm$0.04ms\,$^{-1}$, $\Delta v=2.34\pm$ 0.10 ms\,$^{-1}$ and $C_3=1.75\pm0.03$\,a.u. for figure \ref{FringeFit} (a). The beam in figure \ref{FringeFit} (b) is measured to have $v_g$=16.73$\pm$0.05\,ms$^{-1}$, $\Delta v=2.60\pm$0.16 ms\,$^{-1}$ and $C_3=1.80\pm0.02$\,a.u.. The third data set, figure \ref{FringeFit} (c), gives an atomic beam of $v_g$=19.36$\pm$0.13\,ms$^{-1}$, $\Delta v=3.46\pm$0.32 ms\,$^{-1}$ and $C_3=2.01\pm0.03$\,a.u.. These measurements of $v_g$ are compared with previous measurements in table I. The fixed parameters for fitting the model are $t=160\pm3\,$nm$, \beta = 7.0\pm0.5^{\circ}, s_0=90\pm3\,$nm$, d=257\pm3\,$nm$, L=155.9\pm0.5\,$mm, and $N$=180. The beam profile $A$, from equation \ref{eq:4}, were measured from the beam profile recorded by the MCP detector as shown in figure \ref{SignalProfileRel}.

\begin{table}[htb!]
\begin{tabular}{ |p{1cm}||p{2cm}|p{2cm}|p{2cm}|  }
 \hline
 Data & TOF&Gradient &Fit\\
 \hline
 (a)  & 11.65$\pm$0.49    &13.92$\pm$0.19&   13.90$\pm$0.04\\
 (b)  &   14.04$\pm$0.71  & 16.82$\pm$0.23   &16.73$\pm$0.05\\
 (c)  & 16.67$\pm$0.54 & 19.62$\pm$0.26&  19.36$\pm$0.13\\
 \hline
\end{tabular}
\caption{Comparison of three velocity measurements for each data set shown in figure \ref{FringeFit}. All values are in ms$^{-1}$. TOF refers to the use of the time-of- flight, while gradient is detrmined from the slope of the peak positions shown in figure 14. Fit refers to the fitting of equations 1-5 to the full interferogram.}
\end{table}

The velocity of each beam measured using the gradient and fit are in good agreement, but are approximately 20\% larger than those derived from the TOF. This significant discrepancy highlights the inaccuracy of using TOF to determine the velocity of the beam. The TOF calculation assumes the atoms have an initial velocity close to $v_{TOF}$ and a constant acceleration due to gravity. However, the atoms initial velocity (due to the temperature of the MOT) is much lower and the acceleration is dominated by the short interaction with the push beam laser. After the first collimation slit (or after the push beam pulse), the actual atomic velocity should be higher than the average TOF velocity.  
\subsection{Measuring the van der Waals interaction between the atoms and the grating}
Fitting equations 1-5 also provides a measurement for the VdW coefficient, $C_3$. For the data in figures \ref{FringeFit}(a)-(c), we derive a mean value of 1.84$\pm$0.17\,a.u. from the three fits. There have been no previous attempts to measure $C_3$ for Ar* and Si$_3$N$_4$ as far as we know. This measurement is in good agreement with values derived from spectroscopic data ($C_3$=1.83\,a.u.\cite{Karam2005}).

As shown in figure \ref{dVComb} and \ref{C3Comb}, $\Delta v$ and $C_3$ each change the peak height and signal contrast. In figure \ref{FringeFit}(a), for the interference peaks above the 6th order, the signal noise is too large to determine if the observed envelope is created by the slit function (figure \ref{dVComb}(a)) or the longitudinal velocity spread (figure \ref{C3Comb}(a)).
For a more accurate measurement of velocity and spread, a lower to signal to noise and/or higher contrast peaks would be required. This could be achieved with longer data acquisitions or a narrower longitudinal velocity spread.

\section{Conclusion}
We have characterised a slow, spin-polarised, atomic beam created by resonant acceleration of atoms from a metastable argon MOT. We have shown that matter-wave interferometry is well suited to the measurement of low velocity beams produced by optical forces. As the beam experiences acceleration along its path, we demonstrate that this method is capable of providing precise measurements of the average velocity and velocity distribution, which cannot be obtained from typical TOF measurements. Such measurements will find application in the characterisation of other low velocity beam sources and for future experiments where we aim to measure the weak value of transverse momentum inside a matter-wave interferometer\cite{Morley2016}. Although not compared directly with measurements of beam velocity by Doppler shift measurements, the approach demonstrated here is particularly attractive for low velocity and pulsed beams, where short transit times and Doppler shifts below the laser line width limit its application. Finally, we demonstrate that the slit function, modified by the Van der Waals interactions with the wall, can used to determine the $C_3$ Van der Waals co-efficient for Ar* with the walls of the multi-slit Si$_3$N$_4$ diffraction grating.    

\nocite{*}
\acknowledgements{The authors would like to thank the Franklin Fetzer Foundation for their financial support and Markus Arndt from The University of Vienna for the use of the silicon nitride diffraction grating.}
\bibliography{apssamp}

\providecommand{\noopsort}[1]{}\providecommand{\singleletter}[1]{#1}
\begin{thebibliography}{23}%
\makeatletter
\providecommand \@ifxundefined [1]{%
 \@ifx{#1\undefined}
}%
\providecommand \@ifnum [1]{%
 \ifnum #1\expandafter \@firstoftwo
 \else \expandafter \@secondoftwo
 \fi
}%
\providecommand \@ifx [1]{%
 \ifx #1\expandafter \@firstoftwo
 \else \expandafter \@secondoftwo
 \fi
}%
\providecommand \natexlab [1]{#1}%
\providecommand \enquote  [1]{``#1''}%
\providecommand \bibnamefont  [1]{#1}%
\providecommand \bibfnamefont [1]{#1}%
\providecommand \citenamefont [1]{#1}%
\providecommand \href@noop [0]{\@secondoftwo}%
\providecommand \href [0]{\begingroup \@sanitize@url \@href}%
\providecommand \@href[1]{\@@startlink{#1}\@@href}%
\providecommand \@@href[1]{\endgroup#1\@@endlink}%
\providecommand \@sanitize@url [0]{\catcode `\\12\catcode `\$12\catcode
  `\&12\catcode `\#12\catcode `\^12\catcode `\_12\catcode `\%12\relax}%
\providecommand \@@startlink[1]{}%
\providecommand \@@endlink[0]{}%
\providecommand \url  [0]{\begingroup\@sanitize@url \@url }%
\providecommand \@url [1]{\endgroup\@href {#1}{\urlprefix }}%
\providecommand \urlprefix  [0]{URL }%
\providecommand \Eprint [0]{\href }%
\providecommand \doibase [0]{https://doi.org/}%
\providecommand \selectlanguage [0]{\@gobble}%
\providecommand \bibinfo  [0]{\@secondoftwo}%
\providecommand \bibfield  [0]{\@secondoftwo}%
\providecommand \translation [1]{[#1]}%
\providecommand \BibitemOpen [0]{}%
\providecommand \bibitemStop [0]{}%
\providecommand \bibitemNoStop [0]{.\EOS\space}%
\providecommand \EOS [0]{\spacefactor3000\relax}%
\providecommand \BibitemShut  [1]{\csname bibitem#1\endcsname}%
\let\auto@bib@innerbib\@empty
\bibitem [{\citenamefont {Ramsey}(1985)}]{ramseybeams1}%
  \BibitemOpen
  \bibfield  {author} {\bibinfo {author} {\bibfnamefont {N.}~\bibnamefont
  {Ramsey}},\ }\bibfield  {title} {\bibinfo {title} {Molecular beams},\
  }\href@noop {} {\bibfield  {journal} {\bibinfo  {journal} {Oxford University
  Press}\ } (\bibinfo {year} {1985})}\BibitemShut {NoStop}%
\bibitem [{\citenamefont {Zhelyazkova}\ \emph {et~al.}(2020)\citenamefont
  {Zhelyazkova}, \citenamefont {Martins}, \citenamefont {Agner}, \citenamefont
  {Schmutz},\ and\ \citenamefont {Merkt}}]{PhysRevLett.125.263401}%
  \BibitemOpen
  \bibfield  {author} {\bibinfo {author} {\bibfnamefont {V.}~\bibnamefont
  {Zhelyazkova}}, \bibinfo {author} {\bibfnamefont {F.~B.~V.}\ \bibnamefont
  {Martins}}, \bibinfo {author} {\bibfnamefont {J.~A.}\ \bibnamefont {Agner}},
  \bibinfo {author} {\bibfnamefont {H.}~\bibnamefont {Schmutz}},\ and\ \bibinfo
  {author} {\bibfnamefont {F.}~\bibnamefont {Merkt}},\ }\href@noop {}
  {\bibfield  {journal} {\bibinfo  {journal} {Phys. Rev. Lett.}\ }\textbf
  {\bibinfo {volume} {125}},\ \bibinfo {pages} {263401} (\bibinfo {year}
  {2020})}\BibitemShut {NoStop}%
\bibitem [{\citenamefont {O'Connor}\ \emph {et~al.}(2015)\citenamefont
  {O'Connor}, \citenamefont {Urbain}, \citenamefont {Stützel}, \citenamefont
  {Miller}, \citenamefont {de~Ruette}, \citenamefont {Garrido},\ and\
  \citenamefont {Savin}}]{O_Connor_2015}%
  \BibitemOpen
  \bibfield  {author} {\bibinfo {author} {\bibfnamefont {A.~P.}\ \bibnamefont
  {O'Connor}}, \bibinfo {author} {\bibfnamefont {X.}~\bibnamefont {Urbain}},
  \bibinfo {author} {\bibfnamefont {J.}~\bibnamefont {Stützel}}, \bibinfo
  {author} {\bibfnamefont {K.~A.}\ \bibnamefont {Miller}}, \bibinfo {author}
  {\bibfnamefont {N.}~\bibnamefont {de~Ruette}}, \bibinfo {author}
  {\bibfnamefont {M.}~\bibnamefont {Garrido}},\ and\ \bibinfo {author}
  {\bibfnamefont {D.~W.}\ \bibnamefont {Savin}},\ }\href@noop {} {\bibfield
  {journal} {\bibinfo  {journal} {American Astronomical Society}\ }\textbf
  {\bibinfo {volume} {219}},\ \bibinfo {pages} {6} (\bibinfo {year}
  {2015})}\BibitemShut {NoStop}%
\bibitem [{\citenamefont {Cho}\ and\ \citenamefont
  {Arthur}(1975)}]{CHO1975157}%
  \BibitemOpen
  \bibfield  {author} {\bibinfo {author} {\bibfnamefont {A.}~\bibnamefont
  {Cho}}\ and\ \bibinfo {author} {\bibfnamefont {J.}~\bibnamefont {Arthur}},\
  }\bibfield  {title} {\bibinfo {title} {Molecular beam epitaxy},\ }\href@noop
  {} {\bibfield  {journal} {\bibinfo  {journal} {Progress in Solid State
  Chemistry}\ }\textbf {\bibinfo {volume} {10}},\ \bibinfo {pages} {157}
  (\bibinfo {year} {1975})}\BibitemShut {NoStop}%
\bibitem [{\citenamefont {Jankunas}\ and\ \citenamefont
  {Osterwalder}(2015)}]{beams1}%
  \BibitemOpen
  \bibfield  {author} {\bibinfo {author} {\bibfnamefont {J.}~\bibnamefont
  {Jankunas}}\ and\ \bibinfo {author} {\bibfnamefont {A.}~\bibnamefont
  {Osterwalder}},\ }\bibfield  {title} {\bibinfo {title} {Cold and controlled
  molecular beams: Production and applications},\ }\href@noop {} {\bibfield
  {journal} {\bibinfo  {journal} {Annual Review of Physical Chemistry}\
  }\textbf {\bibinfo {volume} {66}},\ \bibinfo {pages} {241} (\bibinfo {year}
  {2015})}\BibitemShut {NoStop}%
\bibitem [{\citenamefont {Cacciapuoti}\ \emph {et~al.}(2001)\citenamefont
  {Cacciapuoti}, \citenamefont {Castrillo}, \citenamefont {{De Angelis}},\ and\
  \citenamefont {Tino}}]{Cacciapuoti2001}%
  \BibitemOpen
  \bibfield  {author} {\bibinfo {author} {\bibfnamefont {L.}~\bibnamefont
  {Cacciapuoti}}, \bibinfo {author} {\bibfnamefont {A.}~\bibnamefont
  {Castrillo}}, \bibinfo {author} {\bibfnamefont {M.}~\bibnamefont {{De
  Angelis}}},\ and\ \bibinfo {author} {\bibfnamefont {G.~M.}\ \bibnamefont
  {Tino}},\ }\href@noop {} {\bibfield  {journal} {\bibinfo  {journal} {European
  Physical Journal D}\ }\textbf {\bibinfo {volume} {15}},\ \bibinfo {pages}
  {245} (\bibinfo {year} {2001})}\BibitemShut {NoStop}%
\bibitem [{\citenamefont {Wang}\ and\ \citenamefont {Buell}(2003)}]{Wang2003}%
  \BibitemOpen
  \bibfield  {author} {\bibinfo {author} {\bibfnamefont {H.}~\bibnamefont
  {Wang}}\ and\ \bibinfo {author} {\bibfnamefont {W.~F.}\ \bibnamefont
  {Buell}},\ }\href {https://doi.org/10.1364/josab.20.002025} {\bibfield
  {journal} {\bibinfo  {journal} {Journal of the Optical Society of America B}\
  }\textbf {\bibinfo {volume} {20}},\ \bibinfo {pages} {2025} (\bibinfo {year}
  {2003})}\BibitemShut {NoStop}%
\bibitem [{\citenamefont {Maher-McWilliams}\ \emph {et~al.}(2012)\citenamefont
  {Maher-McWilliams}, \citenamefont {Douglas},\ and\ \citenamefont
  {Barker}}]{barker}%
  \BibitemOpen
  \bibfield  {author} {\bibinfo {author} {\bibfnamefont {C.}~\bibnamefont
  {Maher-McWilliams}}, \bibinfo {author} {\bibfnamefont {P.}~\bibnamefont
  {Douglas}},\ and\ \bibinfo {author} {\bibfnamefont {P.~F.}\ \bibnamefont
  {Barker}},\ }\bibfield  {title} {\bibinfo {title} {Laser-driven acceleration
  of neutral particles},\ }\href@noop {} {\bibfield  {journal} {\bibinfo
  {journal} {Nature}\ }\textbf {\bibinfo {volume} {6}},\ \bibinfo {pages} {386}
  (\bibinfo {year} {2012})}\BibitemShut {NoStop}%
\bibitem [{\citenamefont {Lu}\ \emph {et~al.}(1996)\citenamefont {Lu},
  \citenamefont {Corwin}, \citenamefont {Renn}, \citenamefont {Anderson},
  \citenamefont {Cornell},\ and\ \citenamefont {Wieman}}]{Lu2008}%
  \BibitemOpen
  \bibfield  {author} {\bibinfo {author} {\bibfnamefont {Z.~T.}\ \bibnamefont
  {Lu}}, \bibinfo {author} {\bibfnamefont {K.~L.}\ \bibnamefont {Corwin}},
  \bibinfo {author} {\bibfnamefont {M.~J.}\ \bibnamefont {Renn}}, \bibinfo
  {author} {\bibfnamefont {M.~H.}\ \bibnamefont {Anderson}}, \bibinfo {author}
  {\bibfnamefont {E.~A.}\ \bibnamefont {Cornell}},\ and\ \bibinfo {author}
  {\bibfnamefont {C.~E.}\ \bibnamefont {Wieman}},\ }\href
  {https://doi.org/10.1103/PhysRevLett.77.3331} {\bibfield  {journal} {\bibinfo
   {journal} {Phys. Rev. Lett.}\ }\textbf {\bibinfo {volume} {77}},\ \bibinfo
  {pages} {3331} (\bibinfo {year} {1996})}\BibitemShut {NoStop}%
\bibitem [{\citenamefont {Taillandier-Loize}\ \emph {et~al.}(2016)\citenamefont
  {Taillandier-Loize}, \citenamefont {Aljunid}, \citenamefont {Correia},
  \citenamefont {Fabre}, \citenamefont {Perales}, \citenamefont {Tualle},
  \citenamefont {Baudon}, \citenamefont {Ducloy},\ and\ \citenamefont
  {Dutier}}]{Taillandier-Loize2016a}%
  \BibitemOpen
  \bibfield  {author} {\bibinfo {author} {\bibfnamefont {T.}~\bibnamefont
  {Taillandier-Loize}}, \bibinfo {author} {\bibfnamefont {S.~A.}\ \bibnamefont
  {Aljunid}}, \bibinfo {author} {\bibfnamefont {F.}~\bibnamefont {Correia}},
  \bibinfo {author} {\bibfnamefont {N.}~\bibnamefont {Fabre}}, \bibinfo
  {author} {\bibfnamefont {F.}~\bibnamefont {Perales}}, \bibinfo {author}
  {\bibfnamefont {J.~M.}\ \bibnamefont {Tualle}}, \bibinfo {author}
  {\bibfnamefont {J.}~\bibnamefont {Baudon}}, \bibinfo {author} {\bibfnamefont
  {M.}~\bibnamefont {Ducloy}},\ and\ \bibinfo {author} {\bibfnamefont
  {G.}~\bibnamefont {Dutier}},\ }\href@noop {} {\bibfield  {journal} {\bibinfo
  {journal} {Journal of Physics D: Applied Physics}\ }\textbf {\bibinfo
  {volume} {49}} (\bibinfo {year} {2016})}\BibitemShut {NoStop}%
\bibitem [{\citenamefont {Rakonjac}\ \emph {et~al.}(2012)\citenamefont
  {Rakonjac}, \citenamefont {Deb}, \citenamefont {Hoinka}, \citenamefont
  {Hudson}, \citenamefont {Sawyer},\ and\ \citenamefont
  {Kj{\ae}rgaard}}]{Rakonjac:12}%
  \BibitemOpen
  \bibfield  {author} {\bibinfo {author} {\bibfnamefont {A.}~\bibnamefont
  {Rakonjac}}, \bibinfo {author} {\bibfnamefont {A.~B.}\ \bibnamefont {Deb}},
  \bibinfo {author} {\bibfnamefont {S.}~\bibnamefont {Hoinka}}, \bibinfo
  {author} {\bibfnamefont {D.}~\bibnamefont {Hudson}}, \bibinfo {author}
  {\bibfnamefont {B.~J.}\ \bibnamefont {Sawyer}},\ and\ \bibinfo {author}
  {\bibfnamefont {N.}~\bibnamefont {Kj{\ae}rgaard}},\ }\href@noop {} {\bibfield
   {journal} {\bibinfo  {journal} {Opt. Lett.}\ }\textbf {\bibinfo {volume}
  {37}},\ \bibinfo {pages} {1085} (\bibinfo {year} {2012})}\BibitemShut
  {NoStop}%
\bibitem [{\citenamefont {Carey}\ \emph {et~al.}(2018)\citenamefont {Carey},
  \citenamefont {Belal}, \citenamefont {Himsworth}, \citenamefont {Bateman},\
  and\ \citenamefont {Freegarde}}]{Bateman2018}%
  \BibitemOpen
  \bibfield  {author} {\bibinfo {author} {\bibfnamefont {M.}~\bibnamefont
  {Carey}}, \bibinfo {author} {\bibfnamefont {M.}~\bibnamefont {Belal}},
  \bibinfo {author} {\bibfnamefont {M.}~\bibnamefont {Himsworth}}, \bibinfo
  {author} {\bibfnamefont {J.}~\bibnamefont {Bateman}},\ and\ \bibinfo {author}
  {\bibfnamefont {T.}~\bibnamefont {Freegarde}},\ }\href@noop {} {\bibfield
  {journal} {\bibinfo  {journal} {Journal of Modern Optics}\ }\textbf {\bibinfo
  {volume} {65}},\ \bibinfo {pages} {657} (\bibinfo {year} {2018})}\BibitemShut
  {NoStop}%
\bibitem [{\citenamefont {Slowe}\ \emph {et~al.}(2005)\citenamefont {Slowe},
  \citenamefont {Vernac},\ and\ \citenamefont {Hau}}]{Slowe2005}%
  \BibitemOpen
  \bibfield  {author} {\bibinfo {author} {\bibfnamefont {C.}~\bibnamefont
  {Slowe}}, \bibinfo {author} {\bibfnamefont {L.}~\bibnamefont {Vernac}},\ and\
  \bibinfo {author} {\bibfnamefont {L.~V.}\ \bibnamefont {Hau}},\ }\href
  {https://doi.org/10.1063/1.2069651} {\bibfield  {journal} {\bibinfo
  {journal} {Review of Scientific Instruments}\ }\textbf {\bibinfo {volume}
  {76}},\ \bibinfo {pages} {1} (\bibinfo {year} {2005})}\BibitemShut {NoStop}%
\bibitem [{\citenamefont {{Davisson}}\ and\ \citenamefont
  {{Germer}}(1927)}]{davisson}%
  \BibitemOpen
  \bibfield  {author} {\bibinfo {author} {\bibfnamefont {C.}~\bibnamefont
  {{Davisson}}}\ and\ \bibinfo {author} {\bibfnamefont {L.~H.}\ \bibnamefont
  {{Germer}}},\ }\bibfield  {title} {\bibinfo {title} {{The Scattering of
  Electrons by a Single Crystal of Nickel}},\ }\href
  {https://doi.org/10.1038/119558a0} {\bibfield  {journal} {\bibinfo  {journal}
  {\nat}\ }\textbf {\bibinfo {volume} {119}},\ \bibinfo {pages} {558} (\bibinfo
  {year} {1927})}\BibitemShut {NoStop}%
\bibitem [{\citenamefont {Fein}\ \emph {et~al.}(2019)\citenamefont {Fein},
  \citenamefont {Geyer}, \citenamefont {Zwick}, \citenamefont {Kiałka},
  \citenamefont {Pedalino}, \citenamefont {Mayor}, \citenamefont {Gerlich},\
  and\ \citenamefont {Arndt}}]{arndt}%
  \BibitemOpen
  \bibfield  {author} {\bibinfo {author} {\bibfnamefont {Y.~Y.}\ \bibnamefont
  {Fein}}, \bibinfo {author} {\bibfnamefont {P.}~\bibnamefont {Geyer}},
  \bibinfo {author} {\bibfnamefont {P.}~\bibnamefont {Zwick}}, \bibinfo
  {author} {\bibfnamefont {F.}~\bibnamefont {Kiałka}}, \bibinfo {author}
  {\bibfnamefont {S.}~\bibnamefont {Pedalino}}, \bibinfo {author}
  {\bibfnamefont {M.}~\bibnamefont {Mayor}}, \bibinfo {author} {\bibfnamefont
  {S.}~\bibnamefont {Gerlich}},\ and\ \bibinfo {author} {\bibfnamefont
  {M.}~\bibnamefont {Arndt}},\ }\bibfield  {title} {\bibinfo {title} {Quantum
  superposition of molecules beyond 25 kda},\ }\href@noop {} {\bibfield
  {journal} {\bibinfo  {journal} {Nature Physics}\ }\textbf {\bibinfo {volume}
  {15}},\ \bibinfo {pages} {1242} (\bibinfo {year} {2019})}\BibitemShut
  {NoStop}%
\bibitem [{\citenamefont {Rosi}\ \emph {et~al.}(2014)\citenamefont {Rosi},
  \citenamefont {Sorrentino}, \citenamefont {Cacciapuoti},\ and\ \citenamefont
  {et. al}}]{tino}%
  \BibitemOpen
  \bibfield  {author} {\bibinfo {author} {\bibfnamefont {G.}~\bibnamefont
  {Rosi}}, \bibinfo {author} {\bibfnamefont {F.}~\bibnamefont {Sorrentino}},
  \bibinfo {author} {\bibfnamefont {L.}~\bibnamefont {Cacciapuoti}},\ and\
  \bibinfo {author} {\bibnamefont {et. al}},\ }\bibfield  {title} {\bibinfo
  {title} {Precision measurement of the newtonian gravitational constant using
  cold atoms},\ }\href@noop {} {\bibfield  {journal} {\bibinfo  {journal}
  {Nature}\ }\textbf {\bibinfo {volume} {510}},\ \bibinfo {pages} {518}
  (\bibinfo {year} {2014})}\BibitemShut {NoStop}%
\bibitem [{\citenamefont {Peters}\ \emph {et~al.}(2001)\citenamefont {Peters},
  \citenamefont {Chung},\ and\ \citenamefont {Chu}}]{Peters_2001}%
  \BibitemOpen
  \bibfield  {author} {\bibinfo {author} {\bibfnamefont {A.}~\bibnamefont
  {Peters}}, \bibinfo {author} {\bibfnamefont {K.~Y.}\ \bibnamefont {Chung}},\
  and\ \bibinfo {author} {\bibfnamefont {S.}~\bibnamefont {Chu}},\ }\bibfield
  {title} {\bibinfo {title} {High-precision gravity measurements using atom
  interferometry},\ }\href {https://doi.org/10.1088/0026-1394/38/1/4}
  {\bibfield  {journal} {\bibinfo  {journal} {Metrologia}\ }\textbf {\bibinfo
  {volume} {38}},\ \bibinfo {pages} {25} (\bibinfo {year} {2001})}\BibitemShut
  {NoStop}%
\bibitem [{\citenamefont {Berninger}\ \emph {et~al.}(2007)\citenamefont
  {Berninger}, \citenamefont {Stefanov}, \citenamefont {Deachapunya},\ and\
  \citenamefont {Arndt}}]{PhysRevA.76.013607}%
  \BibitemOpen
  \bibfield  {author} {\bibinfo {author} {\bibfnamefont {M.}~\bibnamefont
  {Berninger}}, \bibinfo {author} {\bibfnamefont {A.}~\bibnamefont {Stefanov}},
  \bibinfo {author} {\bibfnamefont {S.}~\bibnamefont {Deachapunya}},\ and\
  \bibinfo {author} {\bibfnamefont {M.}~\bibnamefont {Arndt}},\ }\bibfield
  {title} {\bibinfo {title} {Polarizability measurements of a molecule via a
  near-field matter-wave interferometer},\ }\href
  {https://doi.org/10.1103/PhysRevA.76.013607} {\bibfield  {journal} {\bibinfo
  {journal} {Phys. Rev. A}\ }\textbf {\bibinfo {volume} {76}},\ \bibinfo
  {pages} {013607} (\bibinfo {year} {2007})}\BibitemShut {NoStop}%
\bibitem [{\citenamefont {Nimmrichter}\ \emph {et~al.}(2008)\citenamefont
  {Nimmrichter}, \citenamefont {Hornberger}, \citenamefont {Ulbricht},\ and\
  \citenamefont {Arndt}}]{PhysRevA.78.063607}%
  \BibitemOpen
  \bibfield  {author} {\bibinfo {author} {\bibfnamefont {S.}~\bibnamefont
  {Nimmrichter}}, \bibinfo {author} {\bibfnamefont {K.}~\bibnamefont
  {Hornberger}}, \bibinfo {author} {\bibfnamefont {H.}~\bibnamefont
  {Ulbricht}},\ and\ \bibinfo {author} {\bibfnamefont {M.}~\bibnamefont
  {Arndt}},\ }\bibfield  {title} {\bibinfo {title} {Absolute absorption
  spectroscopy based on molecule interferometry},\ }\href
  {https://doi.org/10.1103/PhysRevA.78.063607} {\bibfield  {journal} {\bibinfo
  {journal} {Phys. Rev. A}\ }\textbf {\bibinfo {volume} {78}},\ \bibinfo
  {pages} {063607} (\bibinfo {year} {2008})}\BibitemShut {NoStop}%
\bibitem [{\citenamefont {Grisenti}\ \emph {et~al.}(1999)\citenamefont
  {Grisenti}, \citenamefont {Sch\"ollkopf}, \citenamefont {Toennies},
  \citenamefont {Hegerfeldt},\ and\ \citenamefont {K\"ohler}}]{Grisenti1999}%
  \BibitemOpen
  \bibfield  {author} {\bibinfo {author} {\bibfnamefont {R.~E.}\ \bibnamefont
  {Grisenti}}, \bibinfo {author} {\bibfnamefont {W.}~\bibnamefont
  {Sch\"ollkopf}}, \bibinfo {author} {\bibfnamefont {J.~P.}\ \bibnamefont
  {Toennies}}, \bibinfo {author} {\bibfnamefont {G.~C.}\ \bibnamefont
  {Hegerfeldt}},\ and\ \bibinfo {author} {\bibfnamefont {T.}~\bibnamefont
  {K\"ohler}},\ }\href {https://doi.org/10.1103/PhysRevLett.83.1755} {\bibfield
   {journal} {\bibinfo  {journal} {Phys. Rev. Lett.}\ }\textbf {\bibinfo
  {volume} {83}},\ \bibinfo {pages} {1755} (\bibinfo {year}
  {1999})}\BibitemShut {NoStop}%
\bibitem [{\citenamefont {Edmunds}\ and\ \citenamefont
  {Barker}(2014)}]{edmunds}%
  \BibitemOpen
  \bibfield  {author} {\bibinfo {author} {\bibfnamefont {P.~D.}\ \bibnamefont
  {Edmunds}}\ and\ \bibinfo {author} {\bibfnamefont {P.~F.}\ \bibnamefont
  {Barker}},\ }\bibfield  {title} {\bibinfo {title} {Trapping cold ground state
  argon atoms},\ }\href {https://doi.org/10.1103/PhysRevLett.113.183001}
  {\bibfield  {journal} {\bibinfo  {journal} {Phys. Rev. Lett.}\ }\textbf
  {\bibinfo {volume} {113}},\ \bibinfo {pages} {183001} (\bibinfo {year}
  {2014})}\BibitemShut {NoStop}%
\bibitem [{\citenamefont {Karam}\ \emph {et~al.}(2005)\citenamefont {Karam},
  \citenamefont {Wipf}, \citenamefont {Grucker}, \citenamefont {Perales},
  \citenamefont {Boustimi}, \citenamefont {Vassilev}, \citenamefont
  {Bocvarski}, \citenamefont {Mainos}, \citenamefont {Baudon},\ and\
  \citenamefont {Robert}}]{Karam2005}%
  \BibitemOpen
  \bibfield  {author} {\bibinfo {author} {\bibfnamefont {J.~C.}\ \bibnamefont
  {Karam}}, \bibinfo {author} {\bibfnamefont {N.}~\bibnamefont {Wipf}},
  \bibinfo {author} {\bibfnamefont {J.}~\bibnamefont {Grucker}}, \bibinfo
  {author} {\bibfnamefont {F.}~\bibnamefont {Perales}}, \bibinfo {author}
  {\bibfnamefont {M.}~\bibnamefont {Boustimi}}, \bibinfo {author}
  {\bibfnamefont {G.}~\bibnamefont {Vassilev}}, \bibinfo {author}
  {\bibfnamefont {V.}~\bibnamefont {Bocvarski}}, \bibinfo {author}
  {\bibfnamefont {C.}~\bibnamefont {Mainos}}, \bibinfo {author} {\bibfnamefont
  {J.}~\bibnamefont {Baudon}},\ and\ \bibinfo {author} {\bibfnamefont
  {J.}~\bibnamefont {Robert}},\ }\href
  {https://doi.org/10.1088/0953-4075/38/15/009} {\bibfield  {journal} {\bibinfo
   {journal} {Journal of Physics B: Atomic, Molecular and Optical Physics}\
  }\textbf {\bibinfo {volume} {38}},\ \bibinfo {pages} {2691} (\bibinfo {year}
  {2005})}\BibitemShut {NoStop}%
\bibitem [{\citenamefont {Morley}\ \emph {et~al.}(2016)\citenamefont {Morley},
  \citenamefont {Edmunds},\ and\ \citenamefont {Barker}}]{Morley2016}%
  \BibitemOpen
  \bibfield  {author} {\bibinfo {author} {\bibfnamefont {J.}~\bibnamefont
  {Morley}}, \bibinfo {author} {\bibfnamefont {P.~D.}\ \bibnamefont
  {Edmunds}},\ and\ \bibinfo {author} {\bibfnamefont {P.~F.}\ \bibnamefont
  {Barker}},\ }\href@noop {} {\bibfield  {journal} {\bibinfo  {journal}
  {Journal of Physics: Conference Series}\ }\textbf {\bibinfo {volume} {701}},\
  \bibinfo {pages} {3} (\bibinfo {year} {2016})}\BibitemShut {NoStop}%
\end{thebibliography}%

\end{document}